\documentclass[11pt]{article}

\usepackage[T1]{fontenc}
\usepackage[utf8]{inputenc}
\usepackage[english]{babel}

\usepackage{amsmath,amssymb}
\usepackage{braket}
\usepackage{physics}
\usepackage{siunitx}
\usepackage{graphicx}
\graphicspath{{Figures_FRET_ArXive/}}
\usepackage{float}

\usepackage[a4paper,margin=2.5cm]{geometry}
\usepackage{booktabs}  

\usepackage[colorlinks=true,
            linkcolor=blue,
            citecolor=blue,
            urlcolor=blue]{hyperref}

\usepackage[numbers]{natbib}

\title{FRET between NV centers in diamond and chlorophyll molecules: a novel resource for multimodal sensing and imaging in plant cells} 

\author{
Sebastian Westrich$^{1}$, Nimba Oshnik$^{1}$, Nina Thiele$^{1}$,\\
Nina Burmeister$^{1}$, Yanis Abdedou$^{1}$, Nikhita Khera$^{1}$,\\
Stefanie J. Müller-Schüssele$^{2}$,\\
Elke Neu$^{1}$\thanks{Corresponding author: nruffing@rptu.de}
}

\begin{document}

\maketitle 

\begin{center}
\small
$^{1}$RPTU University Kaiserslautern--Landau, Department of Physics and State Research Center OPTIMAS,\\
Erwin-Schroedinger-Strasse, 67663 Kaiserslautern, Germany\\[0.3em]
$^{2}$Department of Biology, RPTU University Kaiserslautern--Landau, Germany
\end{center}

\begin{abstract}

This work demonstrates efficient Förster resonance energy transfer (FRET) between ensembles of shallow nitrogen-vacancy (NV) centers located $7\,\text{nm}$ and $9\,\text{nm}$ below a single-crystal diamond surface and a naturally occurring fluorophore, here a mixture of chlorophyll \textit{a} and \textit{b} molecules extracted from \textit{Arabidopsis thaliana}. The broad fluorescence band of NV centers spectrally overlaps with the chlorophyll molecule's absorption, enabling FRET. Consequently, depositing a chlorophyll layer on the diamond surface reduces the NV fluorescence lifetime from approximately $14\,\text{ns}$ to below $4\,\text{ns}$, indicating efficient FRET. Laser-induced photobleaching of chlorophyll restores the unquenched NV lifetime. NV centers located deeper within the diamond ($40\,\text{nm}$ and $72\,\text{nm}$) remain unaffected, confirming that the observed quenching originates from a short range FRET mechanism. The NV ensembles retain their optically detected magnetic resonance (ODMR) contrast while observing FRET, demonstrating preservation of their spin properties. Consequently, these proof-of-principle experiments demonstrate the feasibility of combining FRET-based distance measurements with magnetic sensing using optically readable spins.

\end{abstract}

\section{Introduction}

In fluorescence resonance energy transfer often named Förster resonance energy transfer (FRET), two dipoles, referred to as donor and acceptor,  exchange energy through a non-radiative process \cite{https://doi.org/10.1002/andp.19484370105}. In this process, the donor dipole is excited by absorbing light and subsequently transfers its excitation energy to the acceptor. The acceptor either emits fluorescence or dissipates the energy non-radiatively. In both cases, FRET introduces an additional decay pathway for the donor, thereby reducing its excited-state lifetime and fluorescence intensity. There are two categories of FRET, namely homo-FRET, in which FRET occurs between two identical dipoles, e.g.\ between two identical dye molecules, and hetero-FRET in which non-identical dipoles participate. In the latter case, FRET pairs are feasible in which one dipole is a molecule (dye, fluorescent protein, naturally occurring fluorophore) and the other one is in a solid (quantum dots, point defects in solids). FRET between different types of molecular fluorophores has become a standard resource to measure (intramolecular) distances with high resolution in the sense of a "spectroscopic ruler" \cite{Lakowicz2006,JaresErijman2003,doi:10.1073/pnas.58.2.719,Algar2019}.

Solid-state systems can be particularly stable electrical dipole emitters, making them valuable resources for sensing and imaging. A prominent candidate in this context is the nitrogen-vacancy (NV) center in diamond. Their long-term photostability and nanoscale size enable sensing and imaging with single emitters. In particular, single NV centers have been employed in scanning near-field imaging experiments based on FRET \cite{tisler2013,Schell2014}. NV centers additionally stand out as their fluorescence rate is related to their spin states. A change in the spin state induces a change in the fluorescence rate, thus making the spin states optically readable. Using this property of the NV centers, important measurements in the life sciences have been demonstrated, e.g.\ detection of paramagnetic free radicals in cells connected to neuro-degenerative diseases as well as stress and infections \cite{Schirhagl2022,WU2022102279}. Notably, NV centers in diamond nano-particles have also been shown to undergo hetero-FRET with various synthetic dyes as FRET partners \cite{tisler2011},\cite{https://doi.org/10.1002/adma.200901596},\cite{CHEN2011803}. In our own previous work, we demonstrated hetero-FRET between NV centers, implanted shallowly in a single crystal diamond, and excitons in 2D materials\cite{Nelz2020}. In this model system, we demonstrated that optical spin read-out remains feasible, while the NV centers undergo FRET. With these two prerequisites, in this work, we investigate hetero-FRET between NV centers in a single crystal diamond and naturally occurring fluorophores, namely chlorophyll \textit{a} and \textit{b}, as a novel resource for multimodal sensing and imaging. 
In this work, we thus explore an approach that requires the introduction of only a single foreign fluorophore, namely the NV center, into plant cells. Nanodiamonds containing NV centers can, in principle, be delivered into plant cells using established techniques such as gene guns \cite{Klein1987,Ozyigit2020}. The NV center is particularly attractive for this purpose due to its unique quantum properties, which enable nanoscale magnetic sensing under ambient conditions. By combining the magnetic sensing capabilities of NV centers with the distance sensitivity provided by FRET, this platform offers a promising route for probing intracellular processes with high spatial resolution.

As a proof-of-principle, we build on our previous work \cite{Nelz2020} and employ a model system in which a dense layer of chlorophyll molecules from a plant extract is deposited onto the surface of single-crystal diamond via spin-coating from an ethanolic solution.  In this case, we can control the depth of our NV centers via controlling the energy at which nitrogen ions are implanted into high purity diamond which renders it feasible to investigate the FRET process in detail.

\section{Modeling of the FRET process}

\begin{figure}[!b]
  \includegraphics[width=\linewidth]{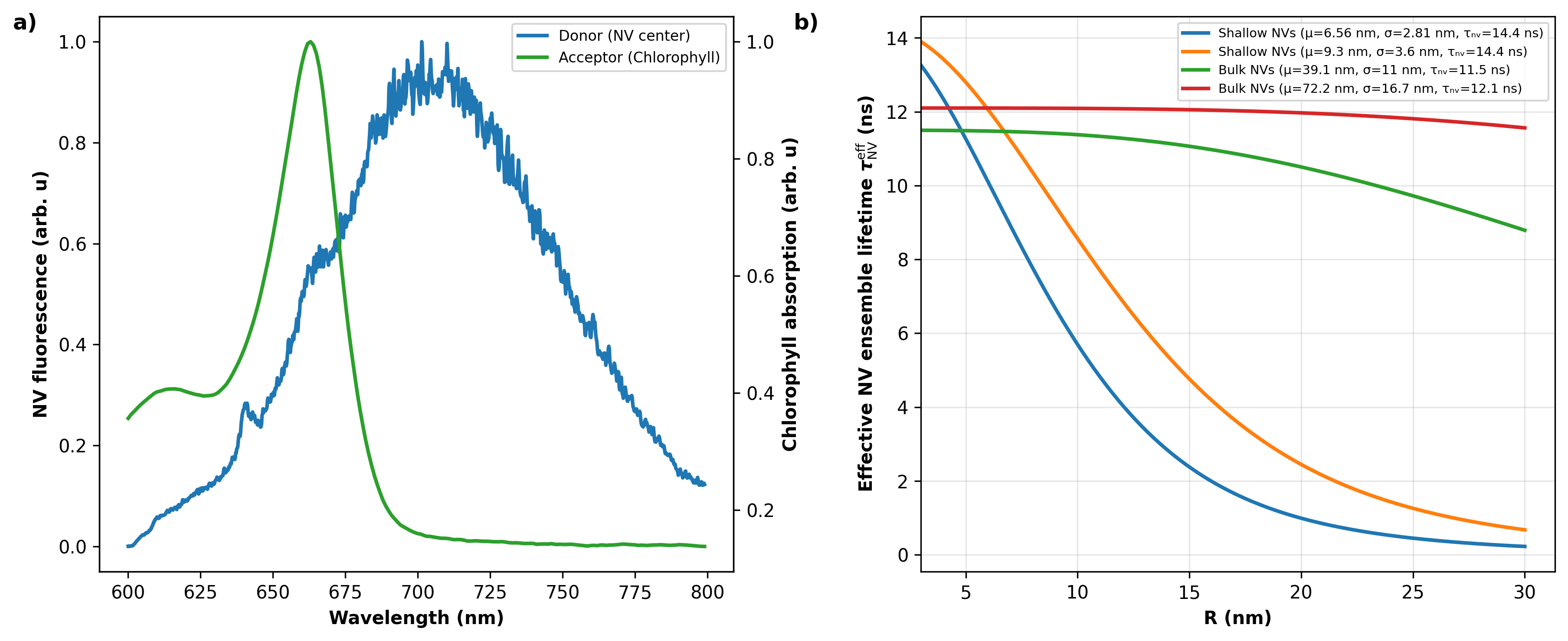}
  \caption{a) Spectral overlap between the NV center emission and the absorption spectrum of the chlorophyll sample. Both spectra were measured experimentally. b) Simulated effective NV ensemble lifetime $\tau_\text{NV}^\text{eff}$ as a function of the effective coupling parameter $R$, calculated using the FRET model described in \textbf{Eq.} \ref{DecyI_Ensemble}.}
  \label{fig:spectraloverlap_FörsterRadius}
\end{figure}

In this work, we demonstrate FRET between a sparse ensemble of shallowly implanted NV centers in single crystal diamond, acting as fluorescent donors, and a naturally occurring acceptor system consisting of a mixture of chlorophyll \textit{a} and \textit{b} molecules extracted from \textit{Arabidopsis thaliana}. We first demonstrate that the NV–chlorophyll pair's spectral properties allow efficient FRET: \textbf{Fig. \ref{fig:spectraloverlap_FörsterRadius} a)} shows the broad fluorescence band of the NV centers spanning 600–800 nm, with the zero-phonon line (ZPL) corresponding to direct optical transition at 637 nm. This band spectrally overlaps with the maximum absorption of the chlorophyll mixture at 664 nm.  The Förster radius $R_0$ between the NV center and the chlorophyll mixture can be calculated using \cite{tisler2011}:

\begin{equation}
R_0 = 0.211(\kappa^2n\textsuperscript{-4}Q\textsubscript{D}J(\lambda))\textsuperscript{1/6},
\label{eq:forster}
\end{equation}

where, $R_0$ is the distance at which the FRET efficiency reaches 50\%. $J(\lambda)$ denotes the spectral overlap between the measured emission spectrum of the NV center and the measured absorption spectrum of the chlorophyll solution. 
The factor $\kappa$ describes the relative orientation of the donor and acceptor dipoles. Here, we assume a random orientation and thus $\kappa^{2} = \frac{2}{3}$. 
The refractive index of the medium is given by $n$, with $n = 2.4$ for diamond. 
The parameter $Q_\text{D}$ corresponds to the quantum efficiency of the donor in the absence of an acceptor. For NV centers shallowly implanted at $4.5 \pm 1\,\text{nm}$ and $8 \pm 2\,\text{nm}$ below the surface, previous work reported quantum efficiencies of $0.7$ and $0.8$, respectively \cite{Radko:16}. From these parameters, we calculate the Förster radius for the NV-chlorophyll pair to be $3.6 \pm 0.1\,\text{nm}$. This result indicates that FRET between the NV center and chlorophyll molecules is feasible for chlorophyll molecules placed on the diamond surface.

The expression in \textbf{Eq. \ref{eq:forster}} and the  $z^{-6}$ distance dependence for FRET between point dipoles describe the energy transfer rate for a single donor acceptor pair. In our experimental model system, however, each NV center in the ensemble faces a dense continuous layer of chlorophyll molecules rather than a single, isolated acceptor. We neglect the thickness of the chlorophyll layer and model it as a two-dimensional acceptor plane. We obtain the total non-radiative rate $\gamma_\text{non-rad}$ for a point donor interacting with an extended two-dimensional acceptor distribution by integrating the pairwise $z^{-6}$ contributions over the acceptor plane. Accordingly, $\gamma_\text{non-rad}$ for this configuration is given by \cite{tisler2013}:

\begin{equation}\label{Modell2}
 \gamma_{non-rad} = \gamma_{rad}\frac{R^4}{z^4}.
\end{equation}

Here, $z$ gives the distance between the donor and the acceptor plane, while $R$ denotes an effective coupling parameter that depends on the Förster radius $R_{0}$ and the (in our case unknown) surface density of acceptor molecules. 
We denote by $\gamma_\text{rad}$ the radiative decay rate of the NV centers and assume equal $\gamma_\text{rad}$ for all NV centers in our ensemble. $\gamma_\text{rad}$ is set equal to ($\tau_\text{{NV}}$)\textsuperscript{-1}, where $\tau_\text{{NV}}$ is the lifetime measured for the different NV ensembles in the absence of FRET. This decay rate implicitly includes all intrinsic non-radiative decay channels present in the bare diamond that are independent of the FRET process. \textbf{Eq. \ref{Modell2}} describes the distance dependence of the FRET induced non-radiative decay. As a result, NV centers located close to the surface experience stronger quenching, leading to shorter fluorescence lifetimes and reduced photon emission rates. However, in our experiment, the detected fluorescence originates from an ensemble of NV centers located at different depths $z$. In each confocal spot, we observe around 10 NV centers. This estimate is based on the ion implantation parameters and the associated NV creation yield (sample preparation see Sec. \ref{sec_sample_prep}). While the detection volume of a confocal microscope extends over several micrometers, the implanted NV ensembles are distributed in depth according to $D(z)$ with a mean and standard deviation on the nanometer scale, resulting in a total depth spread of less than $100\,\text{nm}$. 
Consequently, the measured fluorescence decay curves represent the added emission of all NV centers within the depth distribution $D(z)$.
In principle, this spread in donor–acceptor distances should give rise to a multi-exponential photoluminescence decay. Nevertheless, both, previous reports \cite{Nelz2020} and our own measurements show that the NV ensemble decay can be well described by a single exponential decay corresponding to an averaged lifetime.

To model the measured fluorescence decay of the NV ensemble, we consider the contribution of NV centers located at different depths $z$ below the diamond surface. Each NV center exhibits a mono-exponential decay curve corresponding to its specific donoracceptor distance. These decay curves are weighted by the NV's fluorescence intensity $I(z)$, since NV centers with quenched emission contribute less to the measured signal, and by the depth distribution $D(z)$, which accounts for the geometric distribution of NV centers in the sample. $D(z)$ arises from the ion implantation process and is well described by a Gaussian function characterized by a mean depth $\mu$ and a standard deviation $\sigma$. The overall fluorescence decay of the ensemble is then obtained by integrating over the NV depth profile, leading to 

\begin{equation}
    I(t)_\text{{Ensemble}} = \int_{0}^{\infty} D(z)I_0\frac{\gamma_{r}}{\gamma_{r} + \gamma_{nr}(z)}\exp{(-\frac{t}{\tau(z)})} \,dz \ .
\label{DecyI_Ensemble}
\end{equation}

Using this model, the fluorescence decay of the NV ensemble is well described by a mono-exponential decay with an effective lifetime $\tau_\text{NV}^\text{eff}$, as demonstrated in our previous work \cite{Nelz2020}. We computed $\tau_\text{NV}^\text{eff}$ for the different depth distributions of our samples as a function of $R$ in the range of $3\,\text{nm} < R < 30\,\text{nm}$, \textbf{Fig. \ref{fig:spectraloverlap_FörsterRadius} b)} shows the corresponding graphs. 

\section{Proof-of-principle FRET} 

We first characterize the NV ensembles in four different depths (sample preparation see Sec.\ \ref{sec_sample_prep}, NV depths see \textbf{Tab~\ref{tab:implantation}})), whereas we term the two ensembles at less than 10 nm depth as shallow. All lifetime measurements in this characterization step were carried out on the bare diamond samples before depositing a chlorophyll layer. Shallow NV centers exhibited an average lifetime of $\tau_\text{NV} = 14.4 \pm 0.4\,\text{ns}$, whereas deeper NV centers showed slightly shorter lifetimes of $\tau_\text{NV} = 11.5 \pm 0.5\,\text{ns}$ at $40 \pm 11\,\text{nm} $ depth and $\tau_\text{NV} = 12 \pm 1\,\text{ns}$ at $70 \pm 20\,\text{nm}$. The fluorescence emission of chlorophyll \textit{a} and \textit{b} spans the spectral range from 650 to 800 nm \cite{lichtenthaler1987}, which overlaps with the fluorescence band of NV centers. It is therefore not possible to spectrally separate NV and chlorophyll fluorescence. We measured the fluorescence emission of our chlorophyll sample deposited on silicon substrates and observed a fluorescence maximum at $678\,\text{nm}$, which lies within the range of fluorescence maxima reported in the literature for chlorophyll \textit{a} and \textit{b} \cite{https://doi.org/10.1111/php.13319}. To determine the intrinsic fluorescence lifetime of the chlorophyll mixture, we use a chlorophyll layer on a silicon substrate and excited it using $520\,\text{nm}$ laser light, ensuring direct comparability with the NV-chlorophyll experiments. We obtained an average fluorescence lifetime for chlorophyll of $\tau_\text{Chl} = 0.63 \pm 0.08\,\text{ns}$ indicating that chlorophyll will act as an efficient acceptor and quencher. In addition, $\tau_\text{Chl}$ could be resolved in the measurements on the diamond substrate, where we found an average chlorophyll lifetime of $\tau_\text{Chl} = 0.6 \pm 0.1\,\text{ns}$ extracted from bi-exponential fitting of the decay curves. These results are in good agreement with previously reported in-vivo lifetime measurements on green plants comparable to \textit{Arabidopsis thaliana}, which yielded fluorescence lifetimes of $\tau_\text{Chl} = 0.6\,\text{ns}$ and $\tau_\text{Chl} = 0.7\,\text{ns}$ for chlorophyll \textit{a} aggregates \cite{singhal1969}. 

To investigate stability of the chlorophyll molecules in the film,  we studied the temporal evolution of their fluorescence intensity under continuous-wave (cw) and pulsed laser excitation on both silicon and diamond substrates. Under cw excitation at $520\,\text{nm}$ (laser power $280\,\mu\text{W}$), the fluorescence intensity decreased to approximately $10\,\%$ of its initial value within 15 minutes. Under pulsed excitation at an average laser power of approximately $6\,\mu\text{W}$, a reduction of the fluorescence count rate was also observed, but on a longer timescale of about 45 minutes.  Given that lifetime measurements involve an integration time of $300\,\text{s}$, this slow fluorescence reduction is negligible.  

To investigate FRET between NV centers and chlorophyll molecules, we deposited a chlorophyll layer on the diamond surface by spin-coating (see Sec.~\ref{sec_sample_prep}). The measurement protocol includes the following steps repeated at several positions on the diamond surface:
\begin{itemize}
    \item record fluorescence decay curves using pulsed excitation at $520\,\text{nm}$
    \item bleach chlorophyll molecules using cw excitation ($520\,\text{nm}$, 20 minutes, $280\,\mu\text{W}$ )
    \item repeat measurement of the fluorescence decay curves using pulsed excitation at $520\,\text{nm}$
\end{itemize}
The fluorescence signal, consisting of NV and chlorophyll contributions, was fitted using a two component exponential model convolved with the Gaussian instrument response function (see Sec. \ref{sec_setup}): 
\begin{equation}
\begin{aligned}
I(t) =\;& 
\frac{A_{\text{NV}}}{2}\,
\exp\!\left( \frac{\sigma^2}{2\tau_{\text{NV}}^{2}} - \frac{t - t_0}{\tau_{\text{NV}}} \right)
\left[
1 + \operatorname{erf}\!\left(
\frac{t - t_0 - \frac{\sigma^2}{\tau_{\text{NV}}}}{\sqrt{2}\,\sigma}
\right)
\right] \\
&+ 
\frac{A_{\text{Chl}}}{2}\,
\exp\!\left( \frac{\sigma^2}{2\tau_{\text{Chl}}^{2}} - \frac{t - t_0}{\tau_{\text{Chl}}} \right)
\left[
1 + \operatorname{erf}\!\left(
\frac{t - t_0 - \frac{\sigma^2}{\tau_{\text{Chl}}}}{\sqrt{2}\,\sigma}
\right)
\right]
+ C .
\end{aligned}
\end{equation}

\begin{figure}
  \includegraphics[width=\linewidth]{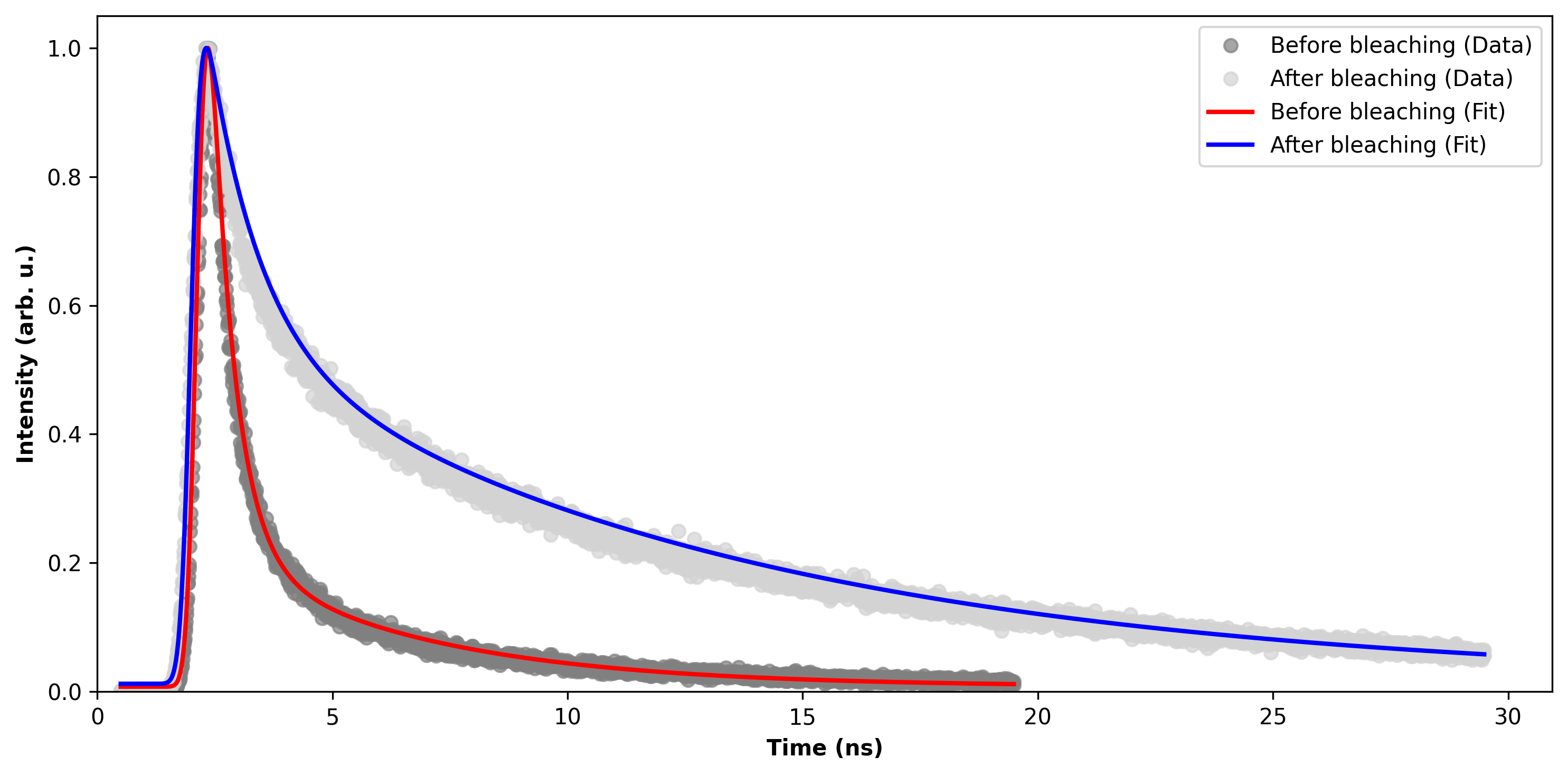}
  \caption{Time-resolved fluorescence measurement of the NV–chlorophyll system. The data correspondsto an NV ensemble located at a depth of $9 \pm 4\,\text{nm}$, excited with pulsed laser light ($\lambda = 520\,\text{nm}$, pulse length $\approx 50\,\text{ps}$). Bleaching of the chlorophyll layer leads to a recovery of the NV lifetime from $4.36 \pm 0.03\,\text{ns}$ to $11.01 \pm 0.05\,\text{ns}$.}
  \label{fig:Lifetimecurves}
\end{figure}

Here, $\tau_{\text{Chl}}$ and $A_{\text{Chl}}$ describe the decay time and amplitude of the chlorophyll fluorescence, respectively, while $\tau_{\text{NV}}$ and $A_{\text{NV}}$ correspond to the fluorescence lifetime and fluorescence amplitude of the NV ensemble. The parameter $\sigma$ denotes the standard deviation of the Gaussian instrument response function  \textbf{Fig. \ref{fig:Lifetimecurves}} illustrates a representative measurement, showing a quenched NV-lifetime of $\tau_\text{{NV, before}} = 4.36 \pm 0.03\,\text{ns}$ and an NV-Lifetime of  $\tau_\text{{NV, after}} = 11.01 \pm 0.05\,\text{ns}$ after bleaching the chlorophyll layer. 

\begin{figure}[H]
  \includegraphics[width=\linewidth]{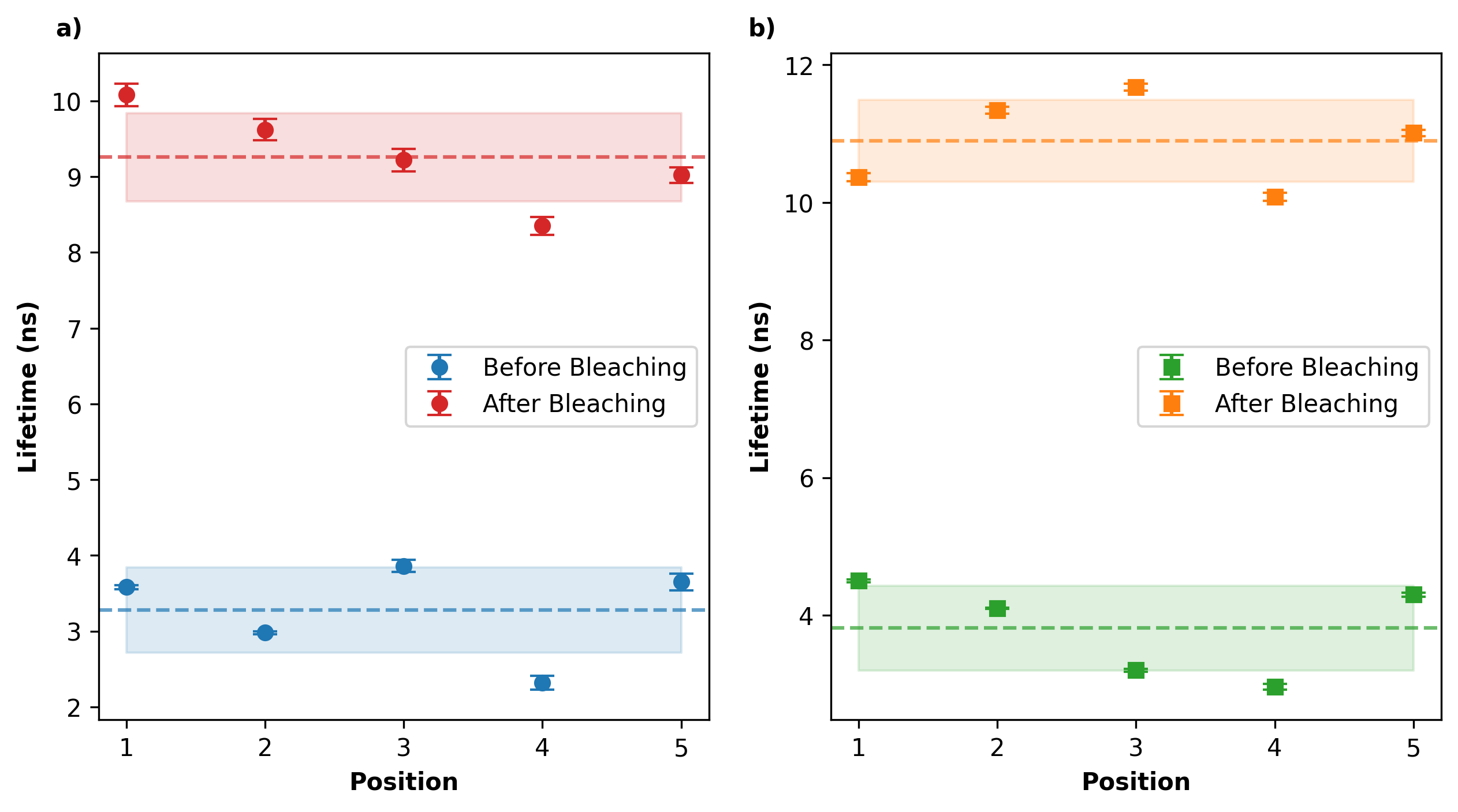}
  \caption{NV fluorescence lifetimes before and after bleaching of the chlorophyll layer. The dashed lines indicate the mean NV lifetimes, and the shaded regions represent one standard deviation ($\pm 1\sigma$). a) NV ensemble with a depth distribution of $7 \pm 3\,\text{nm}$, showing an average NV lifetime of $\tau_\text{NV, before} = 3.3 \pm 0.5\,\text{ns}$ before bleaching and $\tau_\text{NV, after} = 9.3 \pm 0.5\,\text{ns}$ after bleaching of the chlorophyll layer.
  b) NV ensemble with a depth distribution of $9 \pm 4\,\text{nm}$, showing an increase in the average NV lifetime from $\tau_\text{NV, before} = 3.8 \pm 0.6\,\text{ns}$ to $\tau_\text{NV, after} = 10.8 \pm 0.6\,\text{ns}$ following chlorophyll bleaching. In both shallow NV ensembles, chlorophyll efficiently quenches the NV fluorescence, while bleaching restores the original NV lifetime.}
  \label{fig:Lifetime_Results}
\end{figure}

We investigated four diamond samples with NV ensembles in different depths (see \textbf{Tab~\ref{tab:implantation}}). Measurements were performed at several positions across the surface to account for local variations in chlorophyll coverage.
For the NV ensemble with a depth distribution of $7 \pm 3\,\text{nm}$, we observed an average NV lifetime of $\tau_\text{NV, before} = 3.3 \pm 0.5\,\text{ns}$ prior to bleaching. After bleaching the chlorophyll, the NV lifetime increased to $\tau_\text{NV, after} = 9.3 \pm 0.5\,\text{ns}$. A similar trend was found for a second diamond sample with shallow NV centers at $9 \pm 4\,\text{nm}$ depth, where the lifetime increased from $\tau_\text{NV, before} = 3.8 \pm 0.6\,\text{ns}$ to $\tau_\text{NV, after} = 10.8 \pm 0.6\,\text{ns}$. An overview of all measured NV lifetimes before and after chlorophyll bleaching is presented in \textbf{Fig. \ref{fig:Lifetime_Results}}. In both shallow NV ensembles, chlorophyll acted as a strong acceptor and efficiently quenched the NV fluorescence lifetime. Extended laser irradiation bleached the chlorophyll layer and thereby restored the NV lifetime. However, the recovered lifetimes ($\tau_{\text{NV, after}} = 9.3 \pm 0.5\,\text{ns}$ and $10.8 \pm 0.6\,\text{ns}$) do not fully return to the intrinsic value of the untreated NV centers ($14.4 \pm 0.4\,\text{ns}$). We attribute this difference to incomplete bleaching of the chlorophyll layer, in which residual chlorophyll molecules remain on the surface and continue to introduce a weak non-radiative decay channel.
Furthermore, we employed our model of a point donor interacting with an infinite acceptor plane for further analysis. The experimental fluorescence decay curves were fitted using Equation~\ref{DecyI_Ensemble} to extract the effective coupling parameter $R$, which quantifies the donor acceptor interaction strength. To ensure consistency between the measured data and the model, the fitting range was restricted to the time interval dominated by the mono-exponential decay of the NV ensembles. Data from the first nanosecond were excluded, as this region is governed by chlorophyll fluorescence. It should be noted that the effective coupling parameter $R$ obtained here does not correspond to the Förster radius $R_0$ of a single donor-acceptor pair, but rather reflects the combined effect of multiple acceptors and the NV depth distribution in the ensemble. For the ensemble with a depth distribution of $7 \pm 3\,\text{nm}$, we obtained $R = 13 \pm 2\,\text{nm}$, whereas for the ensemble located at $9 \pm 4\,\text{nm}$ depth, we found $R = 17 \pm 2\,\text{nm}$. 

To verify that the observed lifetime reduction is due to FRET, we performed the same measurements on NV ensembles located deeper in the diamond, where the donor–acceptor separation is substantially larger. These measurements were carried out on the same diamond sample, which contains two spatially separated regions implanted with different nitrogen ion energies, resulting in NV depth distributions of $40 \pm 11\,\text{nm}$ and $70 \pm 16\,\text{nm}$. Based on our lifetime simulations, only a minor reduction in the NV ensemble lifetime is expected for these  NV centers. In this case, we measured an average NV lifetime of $\tau_\text{NV, before} = 11.9 \pm 1.1\,\text{ns}$ prior to bleaching and $\tau_\text{NV, after} = 12.1 \pm 0.7\,\text{ns}$ after chlorophyll bleaching. These values agree with the intrinsic lifetimes of the respective NV centers measured prior to chlorophyll deposition ($11.5 \pm  0.5\,\text{ns}$ at $40\,\text{nm}$ and $12 \pm 1\,\text{ns}$ at $70\,\text{nm}$) and show no measurable change upon chlorophyll bleaching. This confirms that NV centers located tens of nanometers below the surface are unaffected, as expected for a short-range FRET mechanism.

\section{Spin properties of NV centers interacting with chlorophyll} 

Moreover, we investigate the electronic spin properties for the NV ensemble (depth: $9.3 \pm 3.6\,\text{nm}$) under the chlorophyll layer. To ensure that chlorophyll does not bleach during the spin measurements, we monitored both the PL count rate and the NV lifetime before and after each optically detected magnetic resonance (ODMR) experiment. Only datasets that showed no significant changes in these parameters were included in the analysis. We used pulsed laser excitation when performing ODMR measurements. Without a chlorophyll layer on top of the diamond surface, we determine an ODMR contrast in the range of 10 to 12 $\%$ in the absence of an external magnetic field. In the presence of the chlorophyll layer, we observed an average ODMR contrast of $4\,\pm 2\,\%$. The reduced ODMR contrast arises from the additional chlorophyll fluorescence, which is not spin-dependent and therefore contributes a background signal. After the chlorophyll was bleached, the ODMR contrast recovered. Our ODMR measurements in the FRET case strongly suggest that spin-based sensing approaches can be combined with FRET-based approaches for NV centers. 

\begin{figure}[H]
  \includegraphics[width=\linewidth]{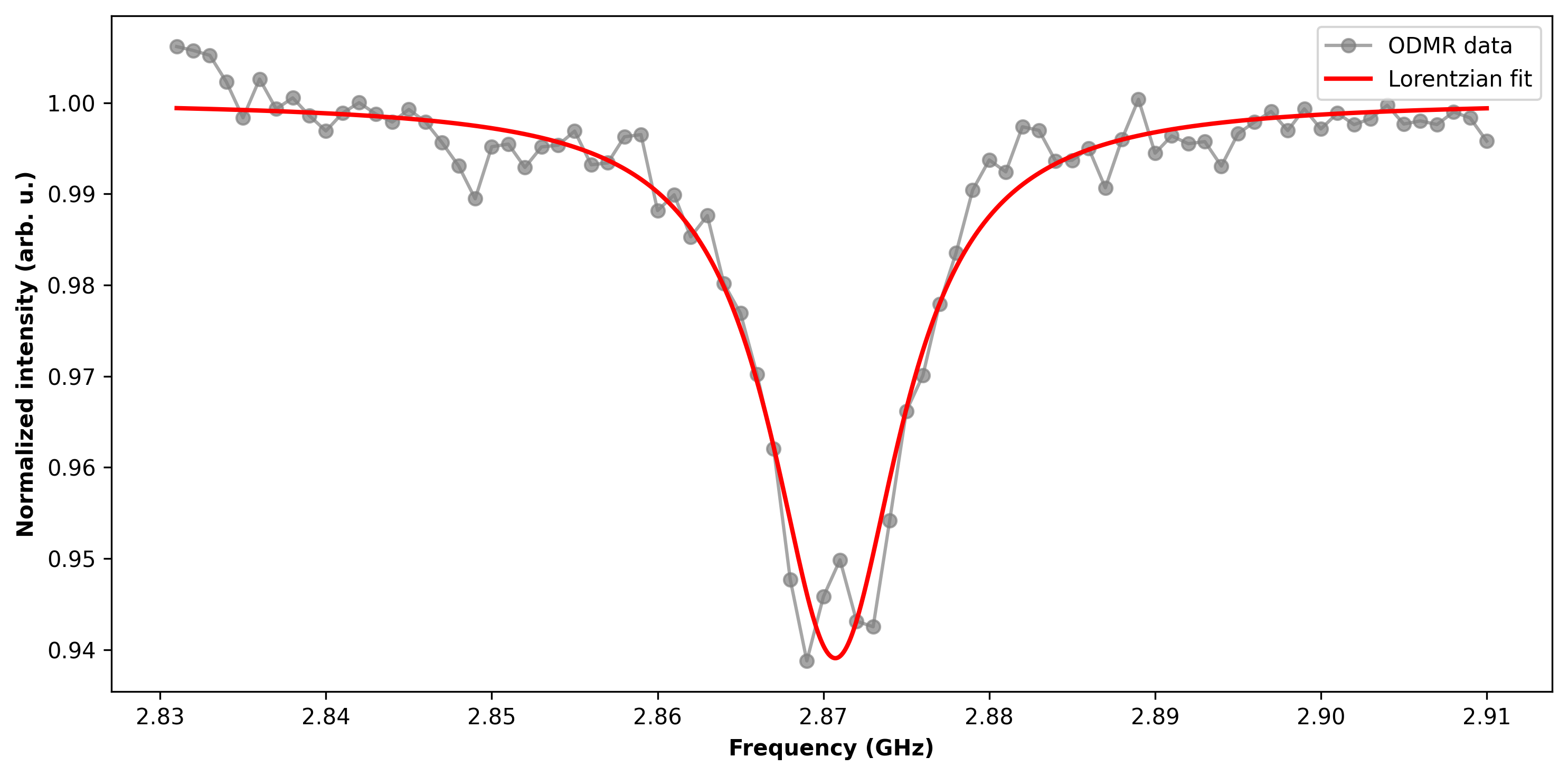}
  \caption{Optically detected magnetic resonance spectrum of the quenched NV ensemble (depth: $9 \pm 4\,\text{nm}$) under the chlorophyll layer in the absence of an external magnetic field. The measurement demonstrates that NV centers maintain their spin-dependent fluorescence contrast even in the presence of efficient energy transfer, highlighting their potential as multifunctional quantum sensors.}
  \label{fig:ZeroFieldODMR_with_Chloro}
\end{figure}

\section{Conclusion}

In conclusion, we have demonstrated FRET between shallow NV centers in single-crystal diamond and a naturally occurring fluorophore, namely a mixture of chlorophyll \textit{a} and \textit{b} extracted from \textit{Arabidopsis thaliana}. The chlorophyll layer acts as a strong energy acceptor, leading to a significant reduction of the NV fluorescence lifetime ($\tau_\text{{NV}} < 4\,\text{ns}$). Bleaching of the chlorophyll restores the unquenched NV lifetime. Importantly, the NV ensembles undergoing FRET retain their optically detected magnetic resonance, demonstrating that their spin properties remain intact during energy transfer. In contrast, NV centers located in the bulk diamond remain unaffected, consistent with the distance dependence of the FRET mechanism.  Our experiments constitute a first step toward using NV centers as sensor in plant cells. While FRET could, e.g.\ determine the proximity to chlorophyll, NV spin measurements can help to detect radical species.

\section{Experimental Section}

\subsection{Experimental Setup \label{sec_setup}} A custom-built confocal scanning microscope (compare Ref. \cite{pandey2022}) was employed to investigate the interaction between NV centers in single-crystal diamond (SCD) and a chlorophyll layer on the diamond surface. For lifetime measurements, we use a tunable pulsed laser (NKT EXW-12, 450--850\,nm, pulse length $\approx 50\,\text{ps}$) equipped with a filter system (NKT SuperK Varia) as the excitation source. A diode laser at 520\,nm (DL nSec, PE 520, Swabian Instruments) with a maximum output power of $40\,\text{mW}$ was utilized to bleach chlorophyll molecules and to perform confocal fluorescence mapping. Excitation light was guided in a single-mode fiber, and a $600\,\text{nm}$ short-pass filter located behind the fiber suppressed background contributions. Sample scanning was performed with a piezoelectric stage (ANSxyz100, Attocube Systems). Excitation and fluorescence light were separated using two dichroic mirrors with a cutoff wavelength of $600\,\text{nm}$ and an additional $600\,\text{nm}$ long-pass filter.  

The collected fluorescence was directed either to a high-efficiency single-photon counting module (Excelitas SPCM-AQRH-14, quantum efficiency $\approx 69\,\%$ at 700\,nm) or to a grating spectrometer (Acton Spectra Pro 2500) equipped with a Pixis 256OE CCD camera. A time-correlated single-photon counting module (PicoQuant PicoHarp 300) was used for time correlated single photon counting to measure the fluorescence decay curves. The instrument response function (IRF) of the setup was determined experimentally by detecting the reflection of the excitation laser pulses from a mirror surface. The resulting IRF is modeled by a Gaussian function with a full width at half maximum (FWHM) of $440\,\text{ps}$. Fluorescence decay curves were fitted with exponential decay functions convoluted with the IRF. For spin manipulation, the setup was equipped with a microwave source (TSG4100A, Tektronix) and a $+45\,\text{dB}$ amplifier (ZHL-16W-43+, Mini-Circuits), while the microwave drive was delivered to the NVs via a $5\,\mu\text{m}$ thick copper wire placed near the sample. 

\subsection{Diamond sample preparation \label{sec_sample_prep}}
 We employed three diamond samples with different NV depth distributions to investigate FRET between NV ensembles and the chlorophyll layer. \linebreak The samples consist of high-purity, (100)-oriented, single-crystal diamonds (Element Six) of electronic grade quality ($[\text{N}] < 5\,\text{ppb}, \; [\text{B}] < 1\,\text{ppb}$). Two of the samples were polished by Almax easyLab and one by Delaware Diamond Knives, achieving a surface roughness of $\text{R}_\text{a} < 1\,\text{nm}$. Following established approaches for creating shallow NV centers \cite{radtke2019, chu2014}, we first remove polishing-induced damage  prior to nitrogen implantation. This is achieved by performing a stress-relief etch using inductively coupled plasma reactive ion etching (ICP-RIE, Oxford Instruments Plasmalab 100), which ensures NV centers in low-strain material. Nitrogen ions were implanted at multiple energies and fluences, under an angle of $7^{\circ}$ to suppress ion channeling. Monte Carlo simulations performed with SRIM \cite{ziegler2010} were used to estimate the corresponding NV center depth distributions. The predicted implantation depths, along with the respective implantation parameters, are summarized in \textbf{Tab~\ref{tab:implantation}}. The two shallow NV ensembles ($<10\,\text{nm}$) were realized on separate diamond samples, whereas the two deeper NV ensembles were created in different regions of a single diamond sample using distinct implantation energies. It should be noted that SRIM simulations typically underestimate experimentally measured implantation depths, as reported in prior studies \cite{PhysRevB.93.045425}.

\begin{table}[ht]
    \centering
    \caption{Nitrogen ion implantation parameters, associated depth distributions, and intrinsic NV lifetimes. \label{tab:implantation}}
    \begin{tabular}{lcccc}
        \hline
        \text{Energy [keV]} & \text{Fluence [cm$^{-2}$]} & \text{Depth [nm]} & \text{Uncertainty [nm]} & \text{Intrinsic NV lifetime [ns]} \\
        \hline
        4  & $4 \times 10^{11}$ & 7  & $\pm 3$  & $14.4 \pm 0.4$ \\
        6  & $3 \times 10^{11}$ & 9  & $\pm 4$  & $14.4 \pm 0.4$ \\
        30 & $5 \times 10^{10}$ & 40 & $\pm 11$ & $11.5 \pm 0.5$ \\
        60 & $3 \times 10^{10}$ & 72 & $\pm 20$ & $12.0 \pm 1.0$ \\
        \hline
    \end{tabular}
\end{table}

After ion implantation, the diamonds were annealed at $800\,^{\circ}\mathrm{C}$ for $2\,\mathrm{hours}$ in vacuum ($<7.8 \times 10^{-7}\,\mathrm{mbar}$) using a custom-built furnace (Edwards T-Station 75, Tectra UHV-compatible Boralectric heater HTR-1001) to form NV centers \cite{chu2014}. To remove residual non-diamond carbon that may form during the annealing process, all diamonds were subsequently cleaned in a tri-acid mixture of nitric acid (65\%), perchloric acid (70\%), and sulfuric acid (96\%) in a 1:1:1 ratio on a hot plate at $500^\circ\mathrm{C}$ for 1\,h. This procedure was followed by sonication in acetone and isopropanol. The same cleaning sequence was also employed to remove chlorophyll residues in-between different measurement cycles. Assuming a typical creation yield on the order of 1\% \cite{Pezzagna_2010}, the implantation parameters result in a sparse NV ensemble with less than 10 NV centers within our confocal microscope's focus. Given the axial resolution in the micron range, the entire NV distribution is effectively contained within the focal depth of our microscope.

\subsection{Chlorophyll extraction}

For the chlorophyll extraction from \textit{Arabidopsis thaliana}, we followed a standard protocol. A volume of 500 $\mu L$ ethanol (95\,\% \, v/v) was added to 10 - 200 mg of plant material in a $2\,\text{mL}$ microcentrifuge tube. The samples were mixed for 5 min in a thermomixer at room temperature and subsequently centrifuged at 14,000 g for 5 min to remove remaining cell debris. A total of $400\,\mu\text{L}$ of the supernatant was then transferred to a fresh microcentrifuge tube. Subsequently, the chlorophyll concentration was determined by absorption spectroscopy. For this purpose, the optical density (OD) of the supernatant was measured at excitation wavelengths of 664 nm, 649 nm, and 470 nm for two different dilution factors. We calculated the concentration of the respective molecules using the method described in \cite{lichtenthaler1987}, and the respective results are shown in \textbf{Tab. \ref{tab:chlorophyll}}.

\begin{table}[!htbp]
\centering
\caption{Concentration of chlorophyll and carotenoids in \textit{Arabidopsis thaliana}.}
\label{tab:chlorophyll}

\resizebox{\textwidth}{!}{%
\begin{tabular}{l S[table-format=3.0] S[table-format=2.2] S[table-format=2.2] S[table-format=3.2] S[table-format=2.2]}
\toprule
Sample & {Dilution Factor} & {Chl a [$\mu$g/mL]} & {Chl b [$\mu$g/mL]} & {Total Chl [$\mu$g/mL]} & {Carotenoids [$\mu$g/mL]} \\
\midrule
\textit{Arabidopsis thaliana} & 10  & 64.63 & 30.88 & 95.50 & 12.67 \\
\textit{Arabidopsis thaliana} & 100 & 59.84 & 46.17 & 106.01 & 7.49 \\
\bottomrule
\end{tabular}%
}
\end{table}

The extracted chlorophyll solution was subsequently used for the preparation of thin films on diamond and silicon substrates.

\subsection{Spin Coating of chlorophyll onto diamond} 

The extracted chlorophyll solution was subsequently used for the preparation of thin films on the diamond substrates by spin coating, which was performed under ambient laboratory conditions. To optimize the process parameters, we first carried out test measurements on silicon substrates using $4\,\mu\text{L}$ of the chlorophyll solution for each spin-coating process. The resulting film quality and surface coverage were then analyzed using atomic force microscopy. AFM [Park XE-70] measurements revealed that the chlorophyll layer is non-uniform and that the molecules tend to aggregate. To evaluate this behavior, we measured the surface roughness of the chlorophyll films. At low spin speeds ($< 1000\,\text{rpm}$), the surface exhibited very high roughness values ($R_a > 100\,\text{nm}$). In contrast, higher spin speeds ($1500 - 3500\,\text{rpm}$) produced smoother films, with an average roughness of $R_a = 20 \pm 8\,\text{nm}$. This aggregation behavior can be attributed to the molecular structure of chlorophylls \textit{a} and \textit{b}, which consist of a chlorin ring and a phytol side chain, a long-chain hydrocarbon. The phytol side chain promotes intermolecular interactions that facilitate molecular clustering on the substrate surface, leading to the observed non-uniform films. In areas without agglomerates, we measured a film thickness of $14\,\pm 3\,\text{nm}$.
For the deposition of the chlorophyll sample onto the diamond, $4\,\mu\text{L}$ of solution was spin-coated at $2000\,\text{rpm}$ for 1 minute. In this case, no AFM roughness analysis was performed to not lose time for the optical measurements. To minimize external influences on the chlorophyll sample with respect to the lifetime measurements, it was necessary to perform the corresponding measurements immediately after coating. The quality and homogeneity of the deposited layer were verified using confocal microscopy. As already observed in preliminary tests with silicon substrates, the formation of agglomerates was also evident on the diamond surface. Nevertheless, chlorophyll molecules were present throughout the substrate. To ensure a reproducible donor acceptor environment, all fluorescence lifetime measurements were performed at positions without visible agglomerates.

\section*{Acknowledgements} 
Funding for this work was provided by the Deutsche Forschungsgemeinschaft (DFG, German Research Foundation) under Grant No. TRR 173–268565370, Spin+X (Project A12). EN acknowledges support from the Quantum-Initiative Rhineland-Palatinate (QUIP). We are grateful to Dr. Oliver Trentmann for experimental assistance.

\section*{Conflict of interest} 
The authors declare no conflict of interest.

\section*{Data Availability Statement} 
The data that support the findings of this study are openly available in Zenodo at he following URL/DOI: 

https://doi.org/10.5281/zenodo.18185936

\newpage

\bibliographystyle{unsrt}
\bibliography{References}

@phdthesis{pandey2022,
  author       = {Pandey, N. O.},
  title        = {Quantum Optimal Control for Quantum Sensing with Nitrogen Vacancy Centers},
  school       = {Technische Universität Kaiserslautern},
  year         = {2022},
  type         = {PhD thesis}
}

@article{radtke2019,
  author    = {Radtke, M. and Render, L. and Nelz, R. and Neu, E.},
  title     = {Plasma Treatments and Photonic Nanostructures for Shallow Nitrogen Vacancy Centers in Diamond},
  journal   = {Optical Materials Express},
  year      = {2019},
  volume    = {9},
  number    = {12},
  pages     = {4716--4733},
  doi       = {10.1364/OME.9.004716}
}

@article{chu2014,
  author    = {Chu, Y. and de Leon, N. P. and Shields, B. J. and Hausmann, B. and Evans, R. and Togan, E. and Burek, M. J. and Markham, M. and Stacey, A. and Zibrov, A. S. and Yacoby, A. and Twitchen, D. J. and Loncar, M. and Park, H. and Maletinsky, P. and Lukin, M. D.},
  title     = {Coherent Optical Transitions in Implanted Nitrogen Vacancy Centers},
  journal   = {Nano Letters},
  year      = {2014},
  volume    = {14},
  number    = {4},
  pages     = {1982--1986},
  doi       = {10.1021/nl404836p}
}

@article{Pezzagna_2010,
doi = {10.1088/1367-2630/12/6/065017},
url = {https://doi.org/10.1088/1367-2630/12/6/065017},
year = {2010},
month = {jun},
volume = {12},
number = {6},
pages = {065017},
author = {Pezzagna, S and Naydenov, B and Jelezko, F and Wrachtrup, J and Meijer, J},
title = {Creation efficiency of nitrogen-vacancy centres in diamond},
journal = {New Journal of Physics},
}

@incollection{lichtenthaler1987,
  author    = {Lichtenthaler, Hartmut K.},
  title     = {Chlorophylls and Carotenoids: Pigments of Photosynthetic Biomembranes},
  booktitle = {Plant Cell Membranes},
  series    = {Methods in Enzymology},
  volume    = {148},
  pages     = {350--382},
  publisher = {Academic Press},
  year      = {1987},
  doi       = {10.1016/0076-6879(87)48036-1},
  url       = {https://www.sciencedirect.com/science/article/pii/0076687987480361}
}

@article{tisler2011,
  author    = {Tisler, Julia and Reuter, Rolf and Lämmle, Anke and Jelezko, Fedor and Balasubramanian, Gopalakrishnan and Hemmer, Philip R. and Reinhard, Friedemann and Wrachtrup, Jörg},
  title     = {Highly Efficient FRET from a Single Nitrogen-Vacancy Center in Nanodiamonds to a Single Organic Molecule},
  journal   = {ACS Nano},
  year      = {2011},
  volume    = {5},
  number    = {10},
  pages     = {7893--7898},
  publisher = {American Chemical Society},
  doi       = {10.1021/nn2021259},
  url       = {https://doi.org/10.1021/nn2021259}
}

@article{tisler2013,
  author    = {Tisler, Julia and Oeckinghaus, Thomas and Stöhr, Rainer J. and Kolesov, Roman and Reuter, Rolf and Reinhard, Friedemann and Wrachtrup, Jörg},
  title     = {Single Defect Center Scanning Near-Field Optical Microscopy on Graphene},
  journal   = {Nano Letters},
  year      = {2013},
  volume    = {13},
  number    = {7},
  pages     = {3152--3156},
  publisher = {American Chemical Society},
  doi       = {10.1021/nl401129m},
  url       = {https://doi.org/10.1021/nl401129m}
}

@article{Nelz2020,
author = {Nelz, Richard and Radtke, Mariusz and Slablab, Abdallah and Xu, Zai-Quan and Kianinia, Mehran and Li, Chi and Bradac, Carlo and Aharonovich, Igor and Neu, Elke},
title = {Near-Field Energy Transfer between a Luminescent 2D Material and Color Centers in Diamond},
journal = {Advanced Quantum Technologies},
volume = {3},
number = {2},
pages = {1900088},
doi = {https://doi.org/10.1002/qute.201900088},
url = {https://advanced.onlinelibrary.wiley.com/doi/abs/10.1002/qute.201900088},
eprint = {https://advanced.onlinelibrary.wiley.com/doi/pdf/10.1002/qute.201900088},
year = {2020}
}

@article{ziegler2010,
  author    = {Ziegler, J. F. and Ziegler, M. D. and Biersack, J. P.},
  title     = {SRIM -- The Stopping and Range of Ions in Matter (2010)},
  journal   = {Nuclear Instruments and Methods in Physics Research Section B: Beam Interactions with Materials and Atoms},
  year      = {2010},
  volume    = {268},
  pages     = {1818--1823},
  doi       = {10.1016/j.nimb.2010.02.091},
  url       = {https://doi.org/10.1016/j.nimb.2010.02.091}
}

@article{PhysRevB.93.045425,
  title = {NMR technique for determining the depth of shallow nitrogen-vacancy centers in diamond},
  author = {Pham, Linh M. and DeVience, Stephen J. and Casola, Francesco and Lovchinsky, Igor and Sushkov, Alexander O. and Bersin, Eric and Lee, Junghyun and Urbach, Elana and Cappellaro, Paola and Park, Hongkun and Yacoby, Amir and Lukin, Mikhail and Walsworth, Ronald L.},
  journal = {Phys. Rev. B},
  volume = {93},
  issue = {4},
  pages = {045425},
  numpages = {12},
  year = {2016},
  month = {Jan},
  publisher = {American Physical Society},
  doi = {10.1103/PhysRevB.93.045425},
  url = {https://link.aps.org/doi/10.1103/PhysRevB.93.045425}
}

@article{singhal1969,
  author    = {Singhal, G. S. and Rabinowitch, E.},
  title     = {Measurement of the Fluorescence Lifetime of Chlorophyll \emph{a} In Vivo},
  journal   = {Biophysical Journal},
  year      = {1969},
  volume    = {9},
  number    = {4},
  pages     = {586--591},
  doi       = {10.1016/S0006-3495(69)86405-2},
  url       = {https://doi.org/10.1016/S0006-3495(69)86405-2}
}

@article{Radko:16,
author = {Ilya P. Radko and Mads Boll and Niels M. Israelsen and Nicole Raatz and Jan Meijer and Fedor Jelezko and Ulrik L. Andersen and Alexander Huck},
journal = {Opt. Express},
number = {24},
pages = {27715--27725},
publisher = {Optica Publishing Group},
title = {Determining the internal quantum efficiency of shallow-implanted nitrogen-vacancy defects in bulk diamond},
volume = {24},
month = {Nov},
year = {2016},
url = {https://opg.optica.org/oe/abstract.cfm?URI=oe-24-24-27715},
doi = {10.1364/OE.24.027715}
}

@book{Lakowicz2006,
  author    = {Joseph R. Lakowicz},
  title     = {Principles of Fluorescence Spectroscopy},
  edition   = {3},
  publisher = {Springer},
  year      = {2006},
  address   = {New York},
  isbn      = {978-0-387-31278-1}
}

@article{JaresErijman2003,
  author    = {Jares-Erijman, Elizabeth A. and Jovin, Thomas M.},
  title     = {FRET imaging},
  journal   = {Nature Biotechnology},
  year      = {2003},
  volume    = {21},
  number    = {11},
  pages     = {1387--1395},
  doi       = {10.1038/nbt896},
  url       = {https://doi.org/10.1038/nbt896},
  issn      = {1546-1696}
}

@article{doi:10.1073/pnas.58.2.719,
author = {L Stryer  and R P Haugland },
title = {Energy transfer: a spectroscopic ruler.},
journal = {Proceedings of the National Academy of Sciences},
volume = {58},
number = {2},
pages = {719-726},
year = {1967},
doi = {10.1073/pnas.58.2.719},
URL = {https://www.pnas.org/doi/abs/10.1073/pnas.58.2.719},
eprint = {https://www.pnas.org/doi/pdf/10.1073/pnas.58.2.719}}

@article{https://doi.org/10.1002/adma.200901596,
author = {Mohan, Nitin and Tzeng, Yan-Kai and Yang, Liling and Chen, Yi-Ying and Hui, Yuen Yung and Fang, Chia-Yi and Chang, Huan-Cheng},
title = {Sub-20-nm Fluorescent Nanodiamonds as Photostable Biolabels and Fluorescence Resonance Energy Transfer Donors},
journal = {Advanced Materials},
volume = {22},
number = {7},
pages = {843-847},
doi = {https://doi.org/10.1002/adma.200901596},
url = {https://advanced.onlinelibrary.wiley.com/doi/abs/10.1002/adma.200901596},
eprint = {https://advanced.onlinelibrary.wiley.com/doi/pdf/10.1002/adma.200901596},
year = {2010}
}

@article{CHEN2011803,
title = {Measuring Förster resonance energy transfer between fluorescent nanodiamonds and near-infrared dyes by acceptor photobleaching},
journal = {Diamond and Related Materials},
volume = {20},
number = {5},
pages = {803-807},
year = {2011},
issn = {0925-9635},
doi = {https://doi.org/10.1016/j.diamond.2011.03.039},
url = {https://www.sciencedirect.com/science/article/pii/S0925963511001245},
author = {Yi-Ying Chen and Hualin Shu and Yung Kuo and Yan-Kai Tzeng and Huan-Chang Chang},
}

@article{Schell2014,
  author    = {Schell, Andreas W. and Engel, Philip and Werra, Julia F. M. and Wolff, Christian and Busch, Kurt and Benson, Oliver},
  title     = {Scanning Single Quantum Emitter Fluorescence Lifetime Imaging: Quantitative Analysis of the Local Density of Photonic States},
  journal   = {Nano Letters},
  year      = {2014},
  volume    = {14},
  number    = {5},
  pages     = {2623--2627},
  doi       = {10.1021/nl500460c},
  url       = {https://doi.org/10.1021/nl500460c},
  issn      = {1530-6984},
  publisher = {American Chemical Society}
}

@article{https://doi.org/10.1002/andp.19484370105,
author = {Förster, Th.},
title = {Zwischenmolekulare Energiewanderung und Fluoreszenz},
journal = {Annalen der Physik},
volume = {437},
number = {1-2},
pages = {55-75},
doi = {https://doi.org/10.1002/andp.19484370105},
url = {https://onlinelibrary.wiley.com/doi/abs/10.1002/andp.19484370105},
eprint = {https://onlinelibrary.wiley.com/doi/pdf/10.1002/andp.19484370105},
year = {1948},
}

@article{Schirhagl2022,
author = {Sigaeva, Alina and Shirzad, Hoda and Martinez, Felipe Perona and Nusantara, Anggrek Citra and Mougios, Nikos and Chipaux, Mayeul and Schirhagl, Romana},
title = {Diamond-Based Nanoscale Quantum Relaxometry for Sensing Free Radical Production in Cells},
journal = {Small},
volume = {18},
number = {44},
pages = {2105750},
doi = {https://doi.org/10.1002/smll.202105750},
url = {https://onlinelibrary.wiley.com/doi/abs/10.1002/smll.202105750},
eprint = {https://onlinelibrary.wiley.com/doi/pdf/10.1002/smll.202105750},
year = {2022}
}

@article{WU2022102279,
title = {Applying NV center-based quantum sensing to study intracellular free radical response upon viral infections},
journal = {Redox Biology},
volume = {52},
pages = {102279},
year = {2022},
issn = {2213-2317},
doi = {https://doi.org/10.1016/j.redox.2022.102279},
url = {https://www.sciencedirect.com/science/article/pii/S2213231722000519},
author = {Kaiqi Wu and Thea A. Vedelaar and Viraj G. Damle and Aryan Morita and Julie Mougnaud and Claudia {Reyes San Martin} and Yue Zhang and Denise P.I. {van der Pol} and Heidi Ende-Metselaar and Izabela Rodenhuis-Zybert and Romana Schirhagl},
}

@article{https://doi.org/10.1111/php.13319,
author = {Taniguchi, Masahiko and Lindsey, Jonathan S.},
title = {Absorption and Fluorescence Spectral Database of Chlorophylls and Analogues},
journal = {Photochemistry and Photobiology},
volume = {97},
number = {1},
pages = {136-165},
doi = {https://doi.org/10.1111/php.13319},
url = {https://onlinelibrary.wiley.com/doi/abs/10.1111/php.13319},
eprint = {https://onlinelibrary.wiley.com/doi/pdf/10.1111/php.13319},
year = {2021}
}

@article{Algar2019,
  author  = {Algar, W. Russ and Hildebrandt, Niko and Vogel, Steven S. and Medintz, Igor L.},
  title   = {FRET as a biomolecular research tool — understanding its potential while avoiding pitfalls},
  journal = {Nature Methods},
  year    = {2019},
  volume  = {16},
  number  = {9},
  pages   = {815--829},
  doi     = {10.1038/s41592-019-0530-8},
  url     = {https://doi.org/10.1038/s41592-019-0530-8},
  issn    = {1548-7105}
}

@article{Klein1987,
  author  = {Klein, T. M. and Wolf, E. D. and Wu, R. and Sanford, J. C.},
  title   = {High-velocity microprojectiles for delivering nucleic acids into living cells},
  journal = {Nature},
  year    = {1987},
  volume  = {327},
  number  = {6117},
  pages   = {70--73},
  doi     = {10.1038/327070a0},
  url     = {https://doi.org/10.1038/327070a0},
  issn    = {1476-4687}
}

@article{Ozyigit2020,
  author  = {Ozyigit, Ibrahim Ilker and Yucebilgili Kurtoglu, Kuaybe},
  title   = {Particle bombardment technology and its applications in plants},
  journal = {Molecular Biology Reports},
  year    = {2020},
  volume  = {47},
  number  = {12},
  pages   = {9831--9847},
  doi     = {10.1007/s11033-020-06001-5},
  url     = {https://doi.org/10.1007/s11033-020-06001-5},
  issn    = {1573-4978}
}

\end{document}